\documentclass[10pt, a4paper]{article}
\usepackage{lrec2022} 
\usepackage{multibib}
\newcites{languageresource}{Language Resources}
\usepackage{graphicx}
\usepackage{tabularx}
\usepackage{placeins}
\usepackage{enumitem}
\usepackage{caption}

\setlist{nosep}
\usepackage{soul}
\usepackage{titlesec}
\titleformat{\section}{\normalfont\large\bfseries\center}{\thesection.}{1em}{}
\titleformat{\subsection}{\normalfont\SmallTitleFont\bfseries\raggedright}{\thesubsection.}{1em}{}
\titleformat{\subsubsection}{\normalfont\normalsize\bfseries\raggedright}{\thesubsubsection.}{1em}{}
\renewcommand\thesection{\arabic{section}}
\renewcommand\thesubsection{\thesection.\arabic{subsection}}
\renewcommand\thesubsubsection{\thesubsection.\arabic{subsubsection}}

\usepackage{epstopdf}
\usepackage[utf8]{inputenc}

\usepackage{hyperref}
\usepackage{xstring}

\usepackage{color}

\usepackage{subcaption}
\usepackage{eurosym}

\title{Political Communities on Twitter: Case Study of the 2022 French Presidential Election \\ \vspace*{.5\baselineskip} }

\name{Hadi Abdine$^{1*}$, Yanzhu Guo$^{1*}$, Virgile Rennard$^{1,2*}$, Michalis Vazirgiannis$^{1,3}$}

\address{$^1$Ecole Polytechnique, $^2$Linagora, $^3$AUEB \\
         \\\{hadi.abdine, yanzhu.guo, virgile.rennard\}@polytechnique.edu\\
         mvazirg@lix.polytechnique.fr}

\abstract{With the significant increase in users on social media platforms, a new means of political campaigning has appeared. Twitter and Facebook are now notable campaigning tools during elections. Indeed, the candidates and their parties now take to the internet to interact and spread their ideas. In this paper, we aim to identify political communities formed on Twitter during the 2022 French presidential election and analyze each respective community. We create a large-scale Twitter dataset containing 1.2 million users and 62.6 million tweets that mention keywords relevant to the election. We perform community detection on a retweet graph of users and propose an in-depth analysis of the stance of each community. Finally, we attempt to detect offensive tweets and automatic bots, comparing across communities in order to gain insight into each candidate's supporter demographics and online campaign strategy.
\newline  }

\begin{document}
\maketitleabstract
\section{Introduction}
\def\thefootnote{*}\footnotetext{These authors contributed equally to this work}
Social media has created a forum for everyone to express themselves, bringing disputes to a wide audience and playing an increasingly crucial part in today's information economy. With the 2008 U.S. presidential election, a relatively new paradigm was observed, where a large part of the political campaign was held on either Facebook or Twitter. The extensive outreach of these platforms has been shown to bring multiple benefits to politicians, such as increases in donations \cite{Petrova2021SocialMA}, or an amplified impact on the politically inattentive youth \cite{10.1111/j.1083-6101.2009.01438.x}. The tremendous amount of data provided by social media platforms gives us insight into the inner workings of the online political horizon.

This paper aims to present an in-depth study on the Twitter landscape of the 2022 French elections. We start by creating a Twitter dataset containing more than 60 million tweets from more than a million users. The tweets are extracted based on keywords related to the election. We use this dataset to build a retweet graph among the users and run a graph-based algorithm for community detection. By analyzing the top hashtags and word clouds of tweets posted by users of each community, we are able to interpret which candidate they each support. We go on to visualize the geographical distribution of each candidate's online supporters across different regions, making comparisons between communities. Eventually, we perform offensiveness detection and bot detection in all the communities. The detection of offensive tweets reveals that supporters of certain candidates are more likely to post offensive contents. The results of bot detection also indicates that there are higher levels of bot activities in certain online communities than in others. However, we would like to emphasize that the results of both offensiveness detection and bot detection are produced by automatic classification models reflecting patterns of the datasets they were trained on. Such models are subject to various limitations and by no means reflect our personal opinions.

The rest of this paper is organized as follows. Section 2 provides an overview of the related work. Section 3 describes our the dataset we use and how we collected it. Section 4 supplies a detailed description of the graph-based communities. Section 5 detects offensive tweets in each community while section 6 studies the use of automated bot accounts. Finally, section 7 summarizes our research and presents potential future work.
    
\section{Related Work}
Early work on Twitter analysis of political elections dates back to when Twitter was founded. An example is a study on the 2008 U.S. presidential election \cite{diakopoulos2010characterizing} where the debate performance of presidential candidates is characterized by aggregated Twitter sentiment. This initiated a branch of research centered around monitoring online public reactions during election periods. Relevant studies have been carried out on a wide range of elections in different countries, including the 2012 South Korea presidential election \cite{bae2013analysis}, the 2013 German parliamentary election \cite{RILL201424}, and the 2017 UK general election \cite{10.1145/3339909}.

Another popular branch of research aims at forecasting election results based on Twitter data. For example, a study on the 2009 German federal election \cite{tumasjan2010predicting} claims that the respective shares of Twitter volume can accurately reflect the distribution of electoral votes for the six main parties. However, another study on the 2011 Singapore presidential election using Twitter sentiment succeeded in picking out the top two candidates but failed to predict the final ranking \cite{choy2011sentiment}. It is generally agreed upon that Twitter analysis for election outcome prediction cannot substitute traditional polling approaches \cite{bermingham2011using}, and that explainable models should accompany the predictive results \cite{gayo2011limits}.

More recently, attention has been drawn to the diffusion of misinformation and toxicity during online campaigns. Relevant topics include the detection of fake news \cite{cinelli2020limited}, social bots \cite{pastor2020spotting}, political trolls \cite{badawy2018analyzing} as well as hate speech \cite{siegel2021trumping} and offensive language \cite{grimminger2021hate}. Our work is closely related to this field of study while also drawing upon graph-based analysis of Twitter network structures \cite{radicioni2021analysing}.
    
\begin{figure*}[hp!]
	\centering
	\subfloat[Macron]{
	\includegraphics[width=0.45\textwidth]{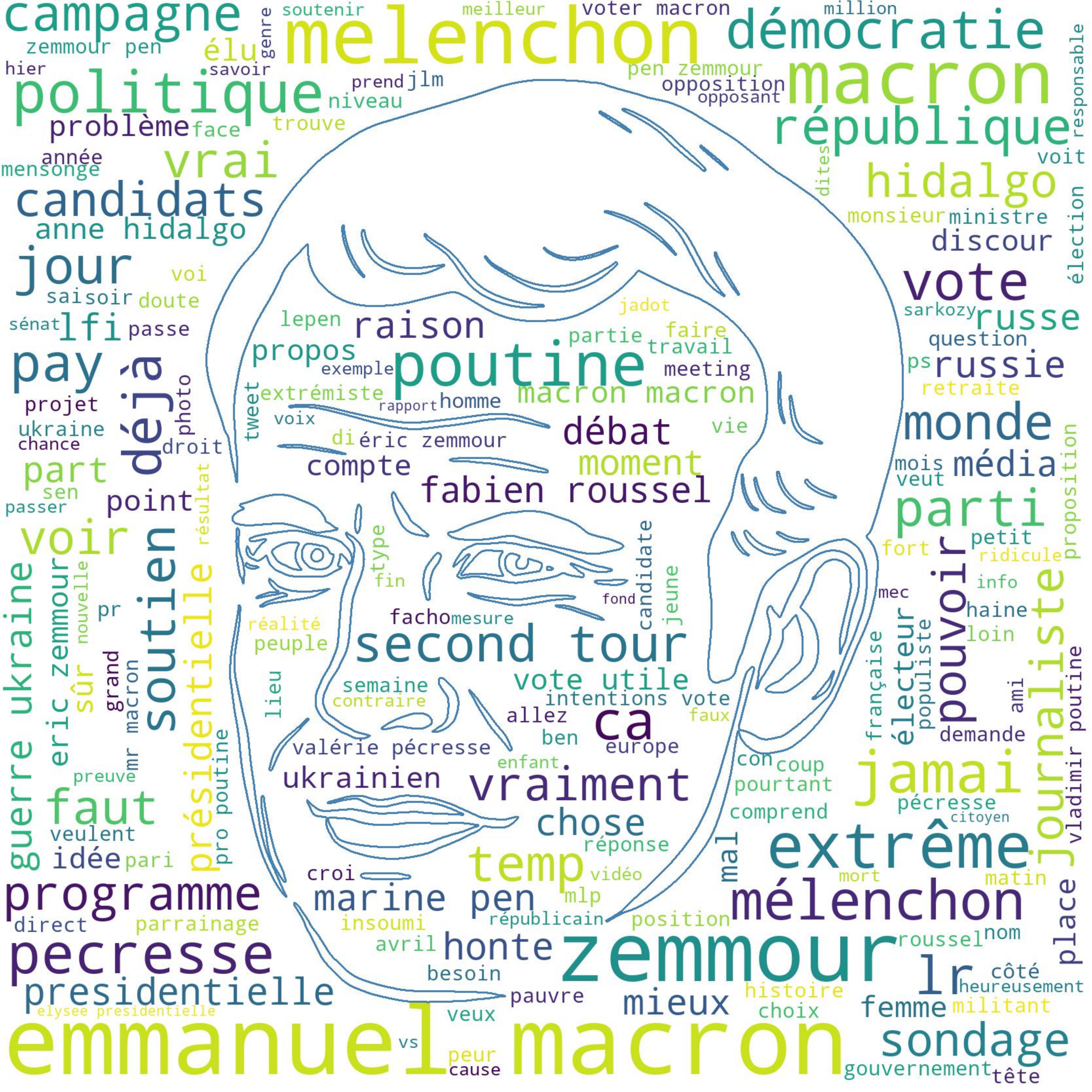}
	}
	\subfloat[Pécresse]{
	\includegraphics[width=0.45\textwidth]{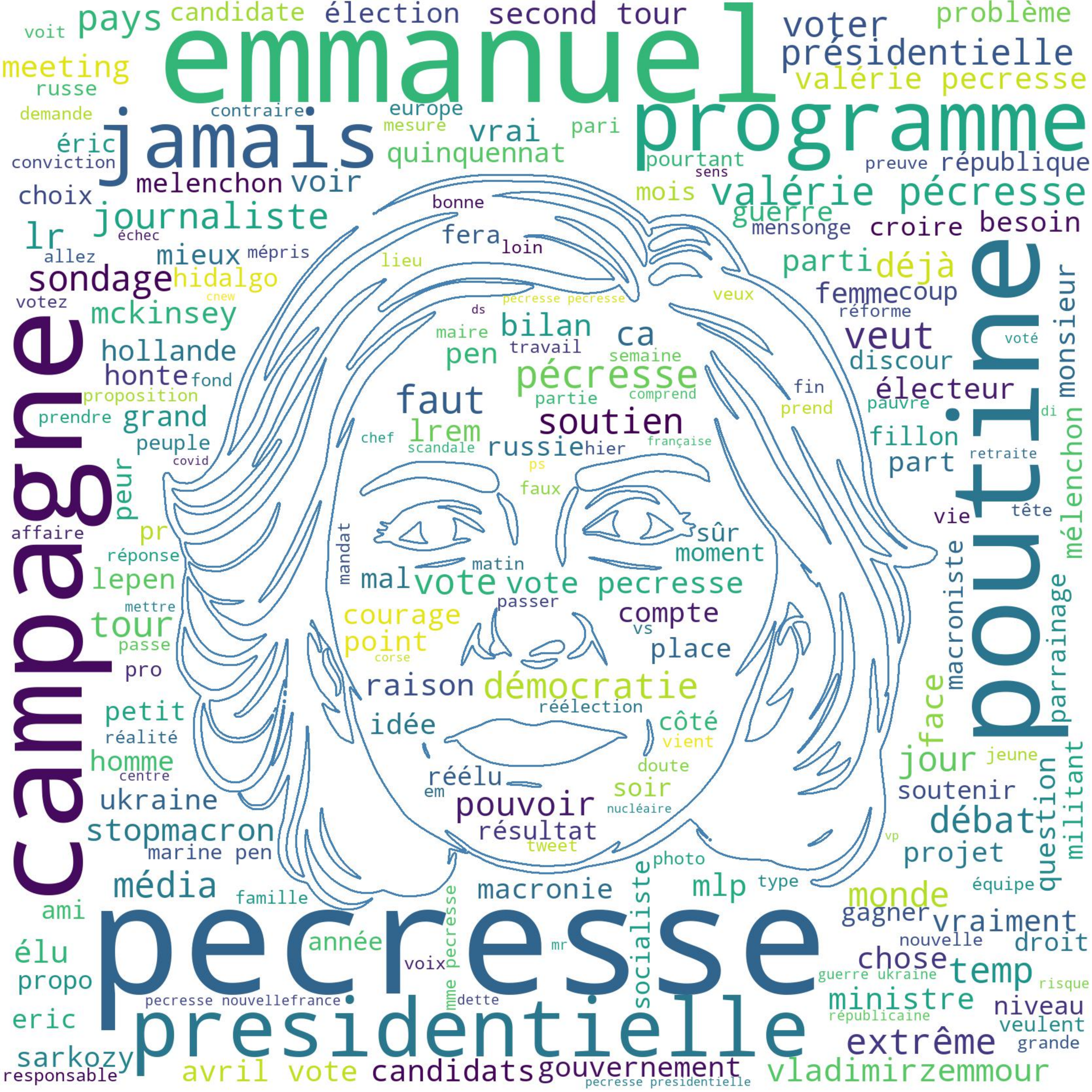}
	}\\
	\subfloat[Zemmour]{
	\includegraphics[width=0.45\textwidth]{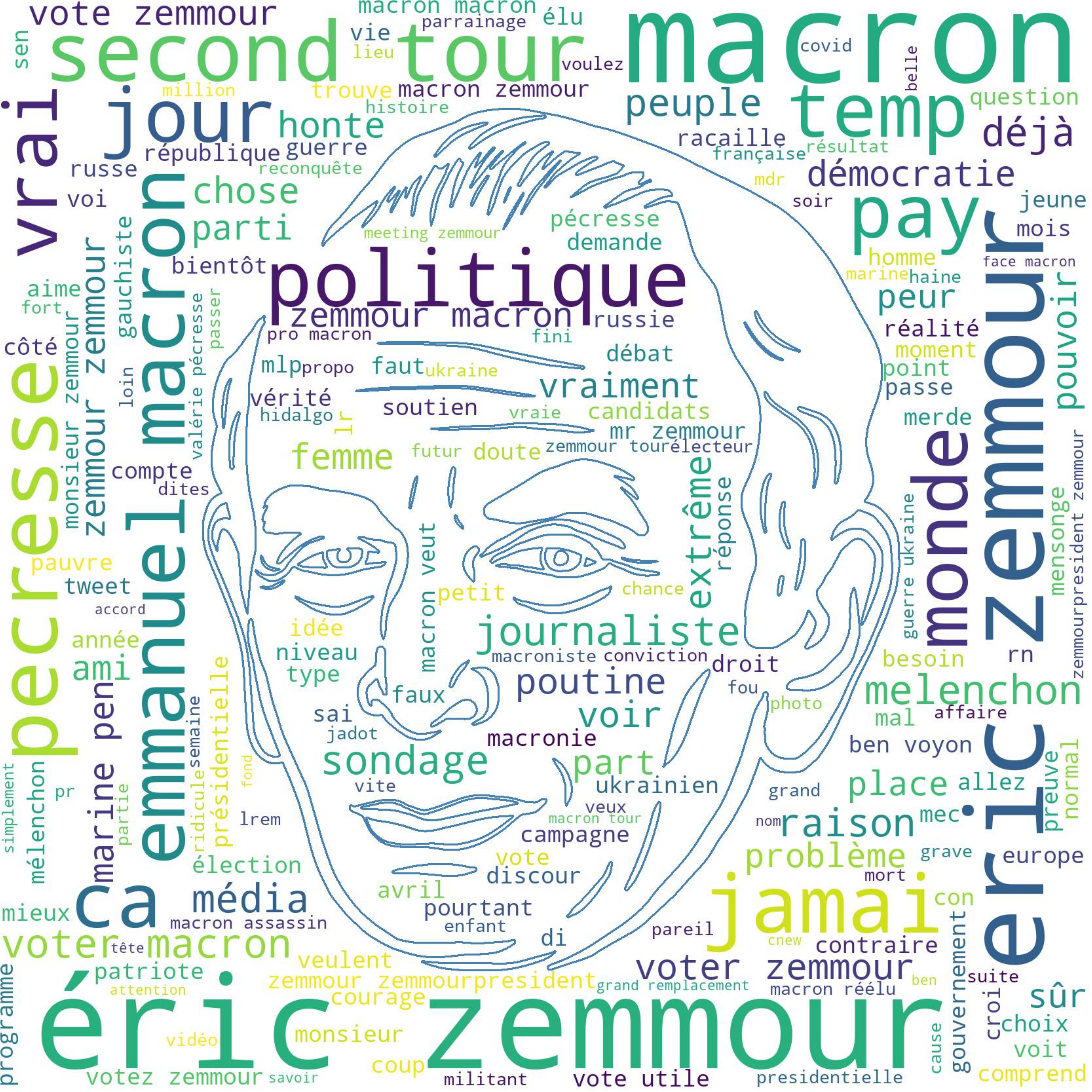}
	}
	\subfloat[Mélenchon]{
	\includegraphics[width=0.45\textwidth]{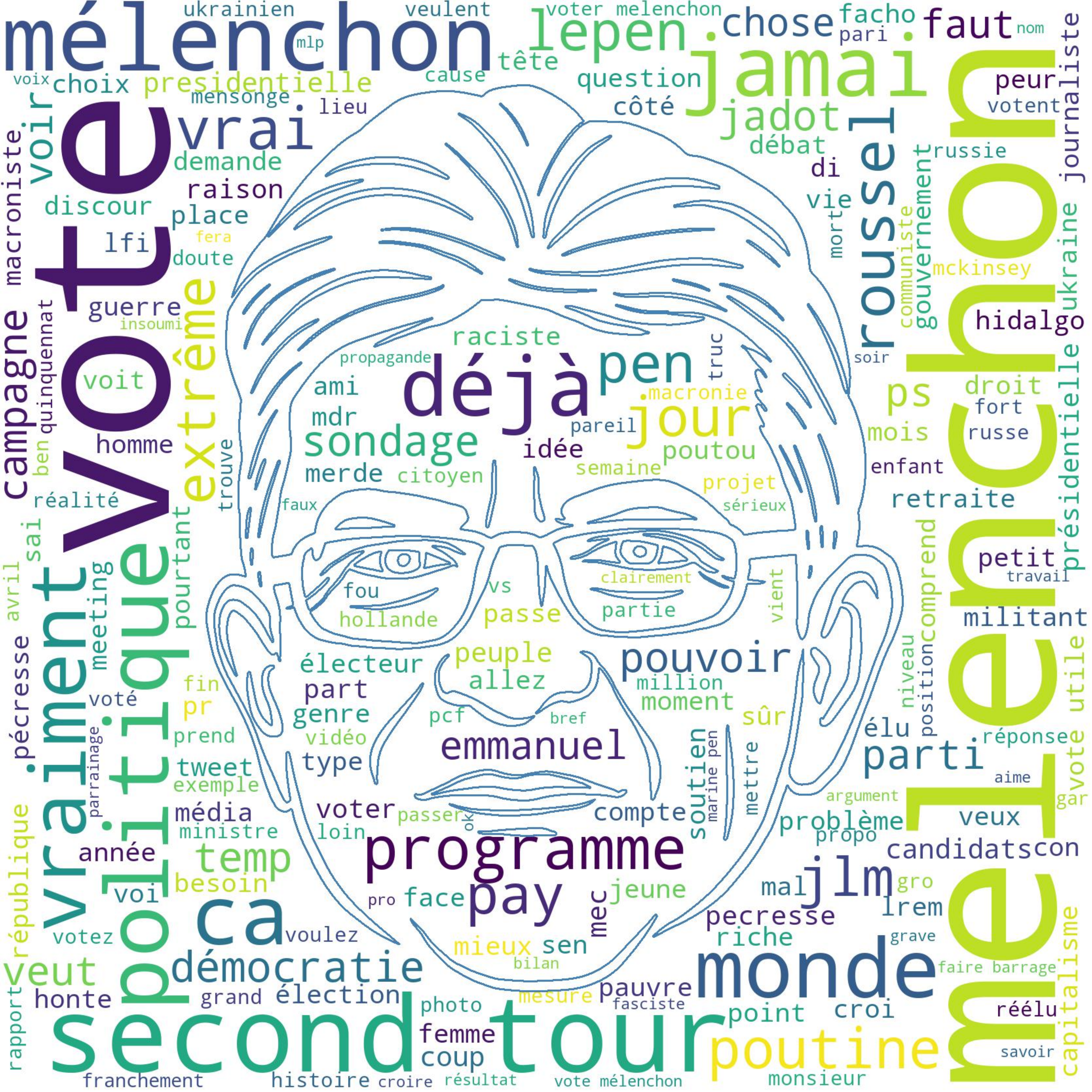}
	}\\
	\subfloat[Le Pen]{
	\includegraphics[width=0.45\textwidth]{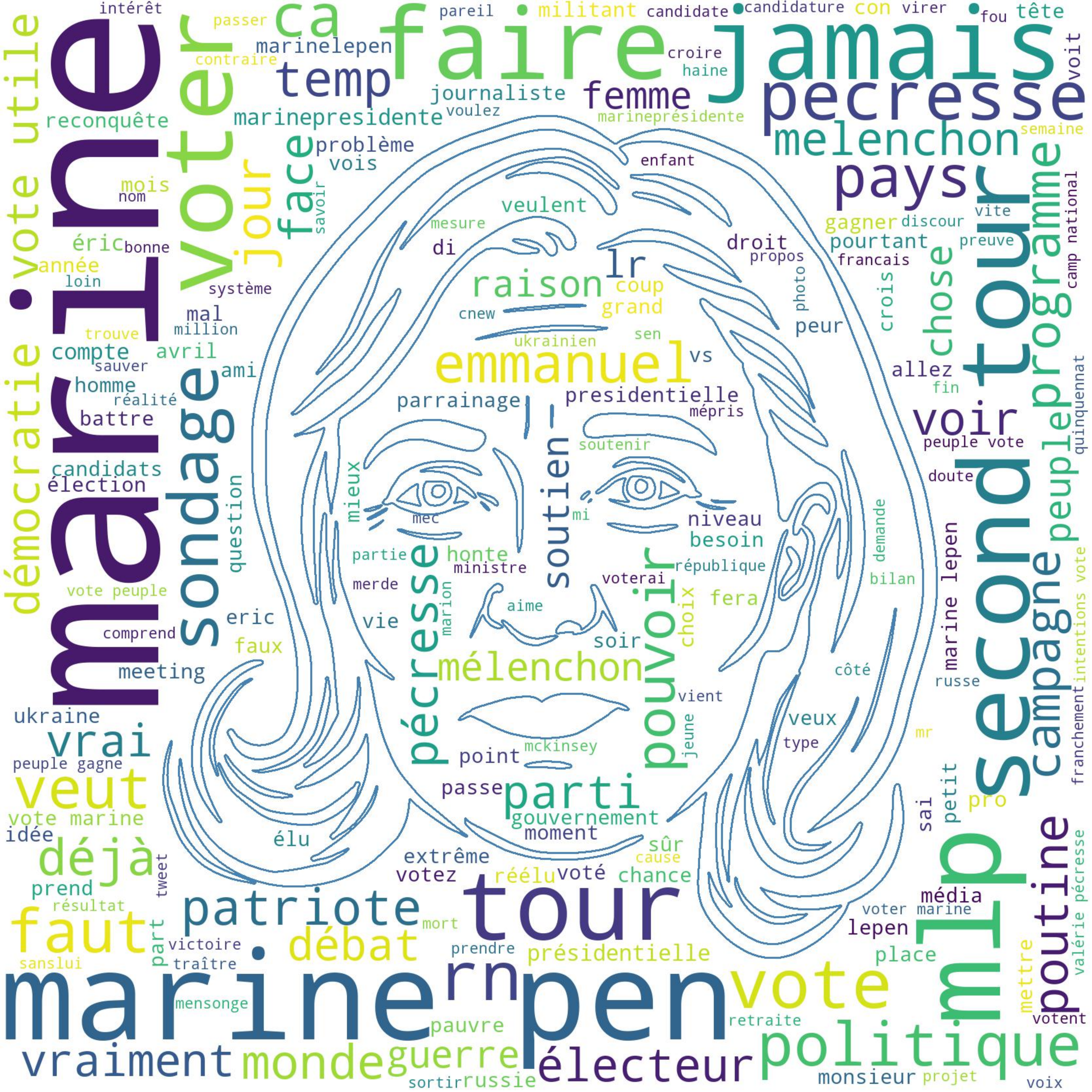}
	}
	\subfloat[Jadot]{
	\includegraphics[width=0.45\textwidth]{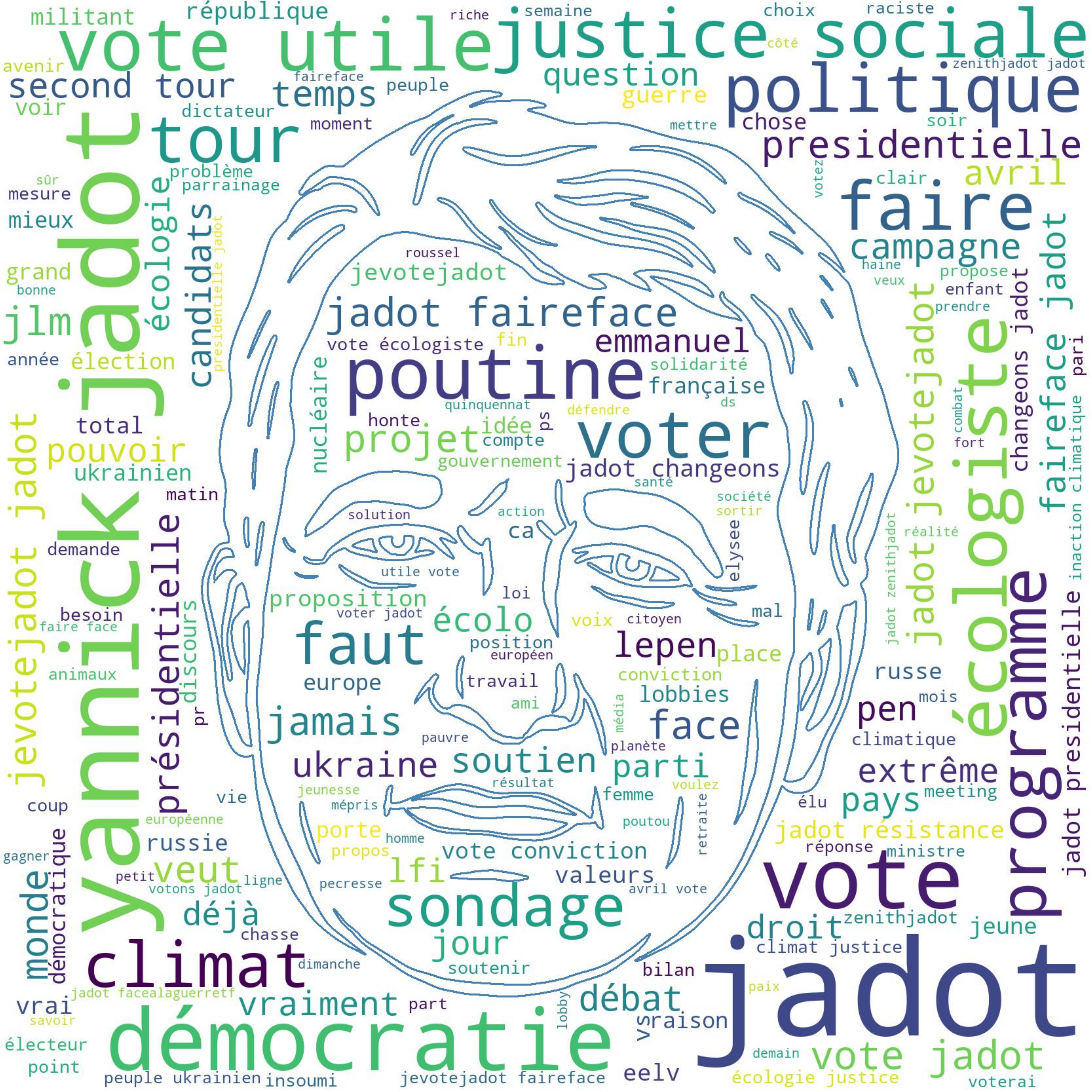}
	}\\
	\caption{Word clouds generated from tweets posted by users in each community.}
	\label{wordcloud}
\end{figure*}

\begin{figure*}[ht!]
	\centering
	\subfloat[Macron]{
	\includegraphics[width=0.45\textwidth]{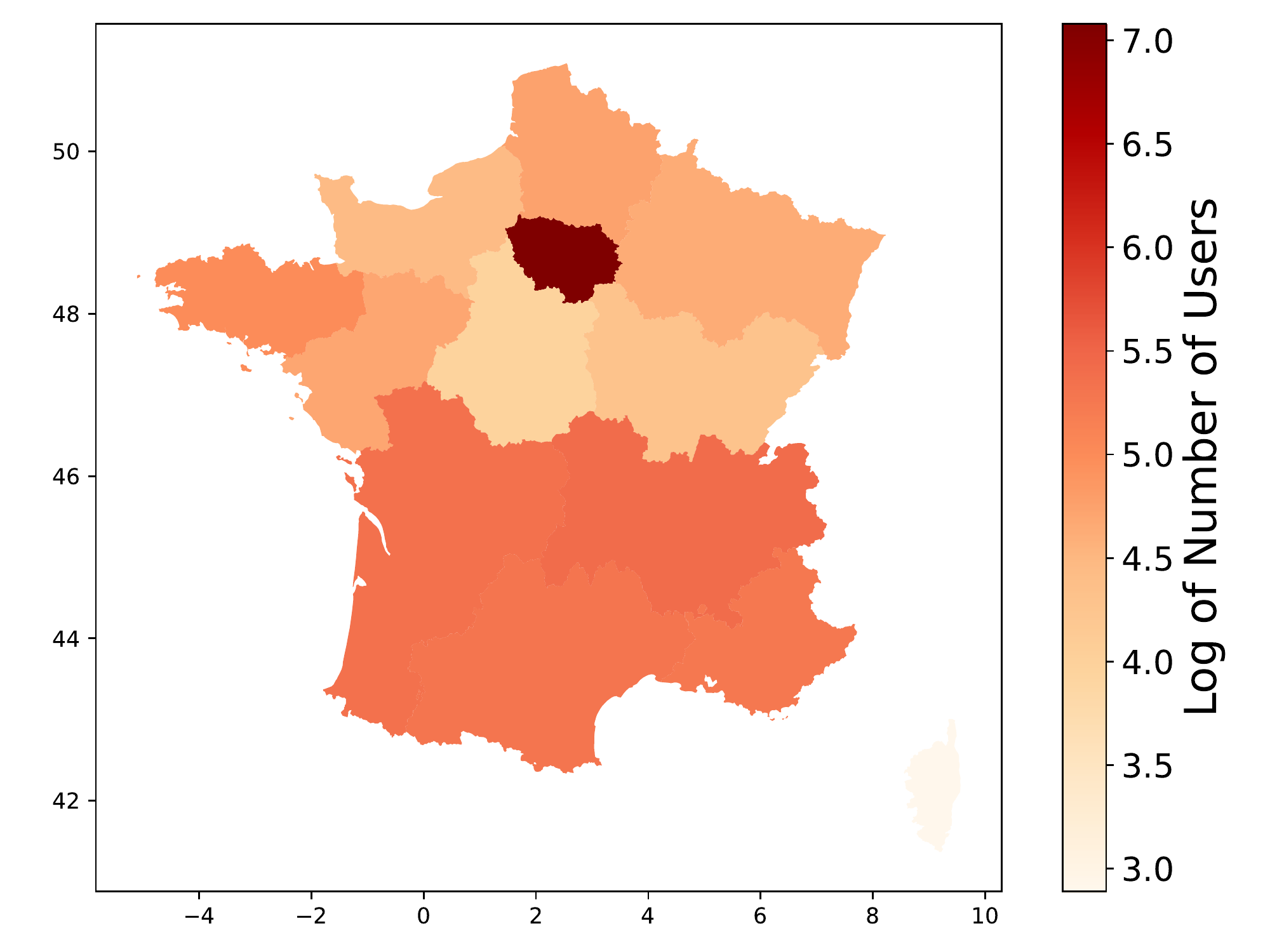}
	}
	\subfloat[Pécresse]{
	\includegraphics[width=0.45\textwidth]{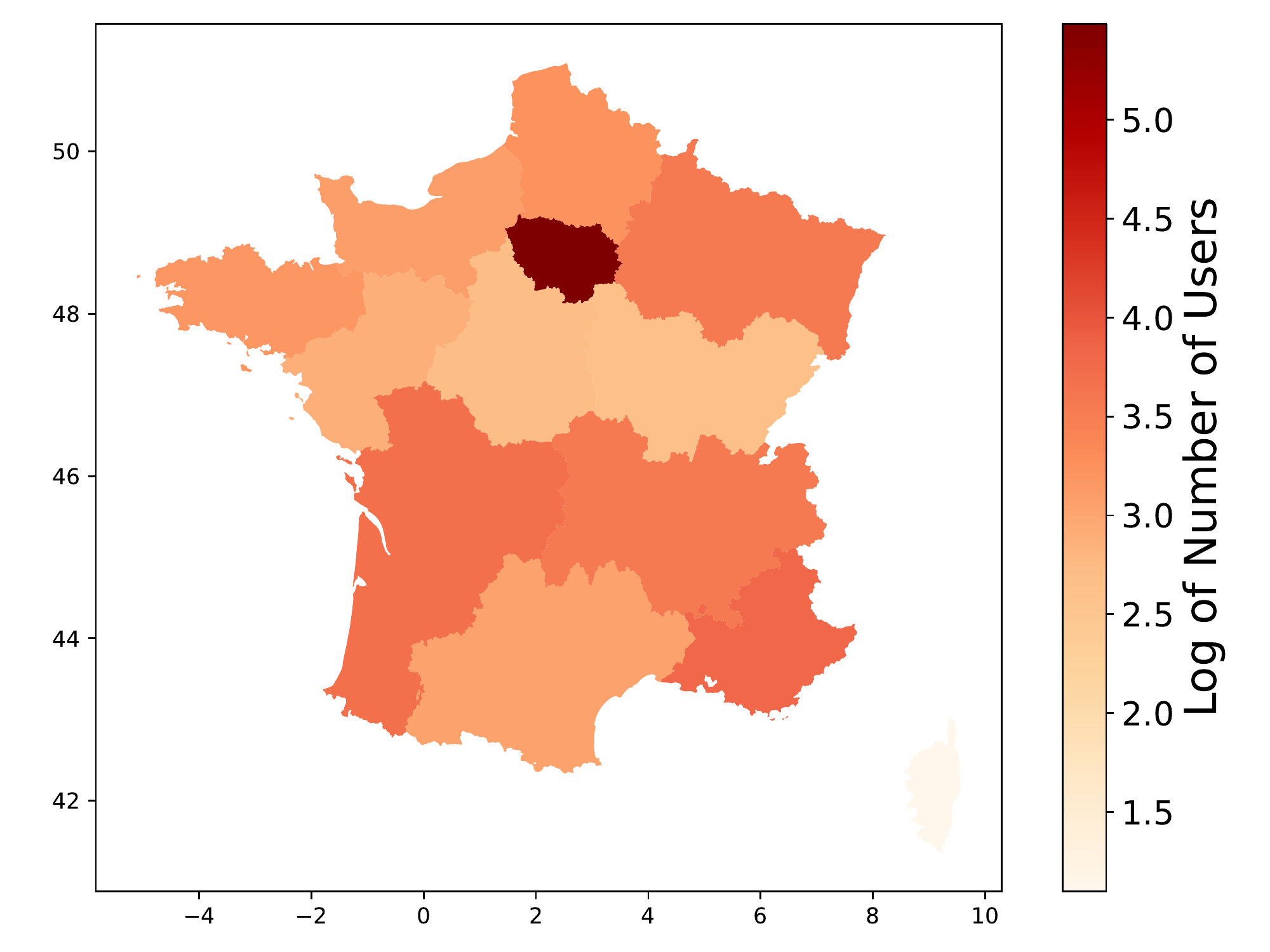}
	}\\
	\subfloat[Zemmour]{
	\includegraphics[width=0.45\textwidth]{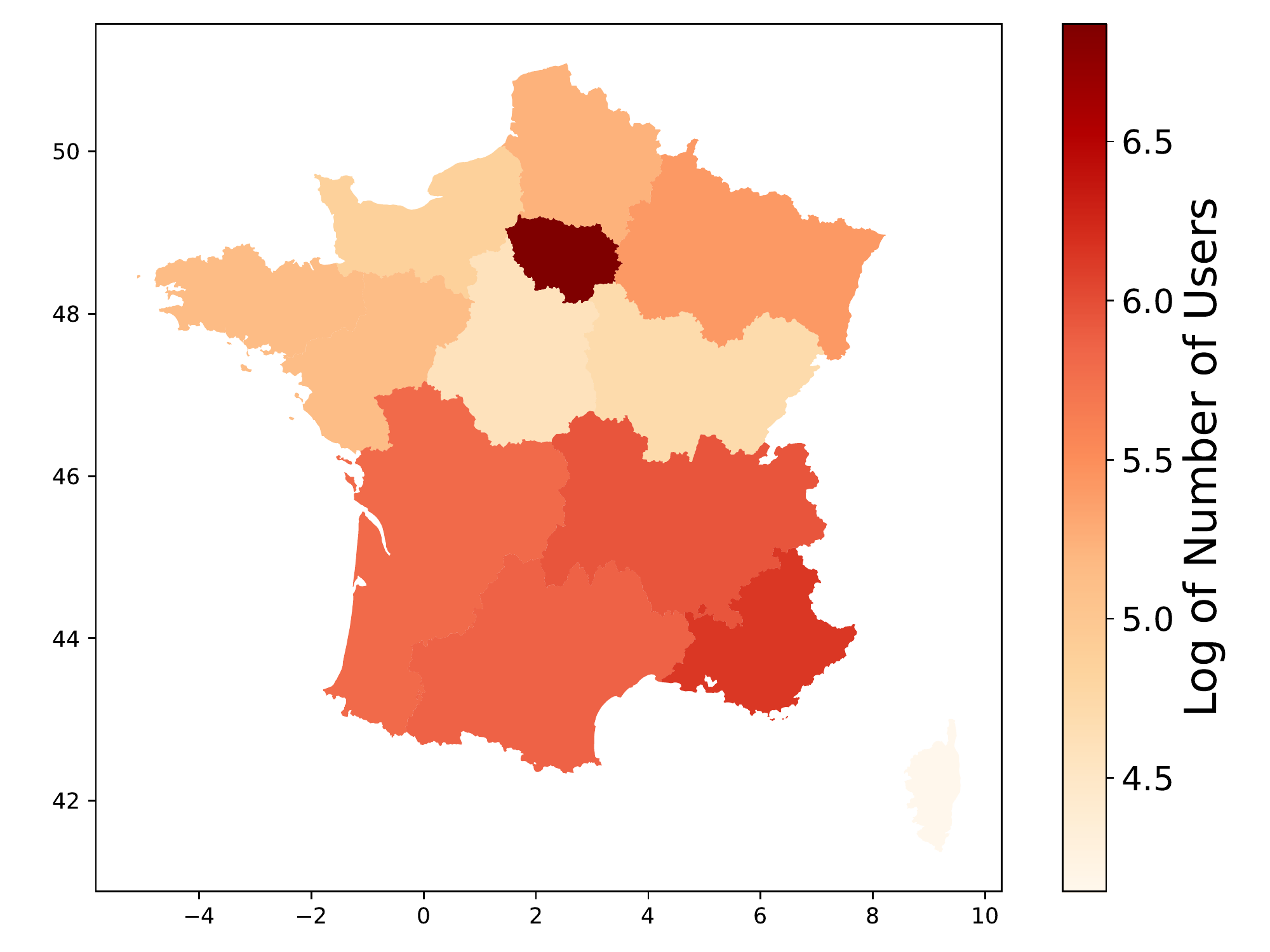}
	}
	\subfloat[Mélenchon]{
	\includegraphics[width=0.45\textwidth]{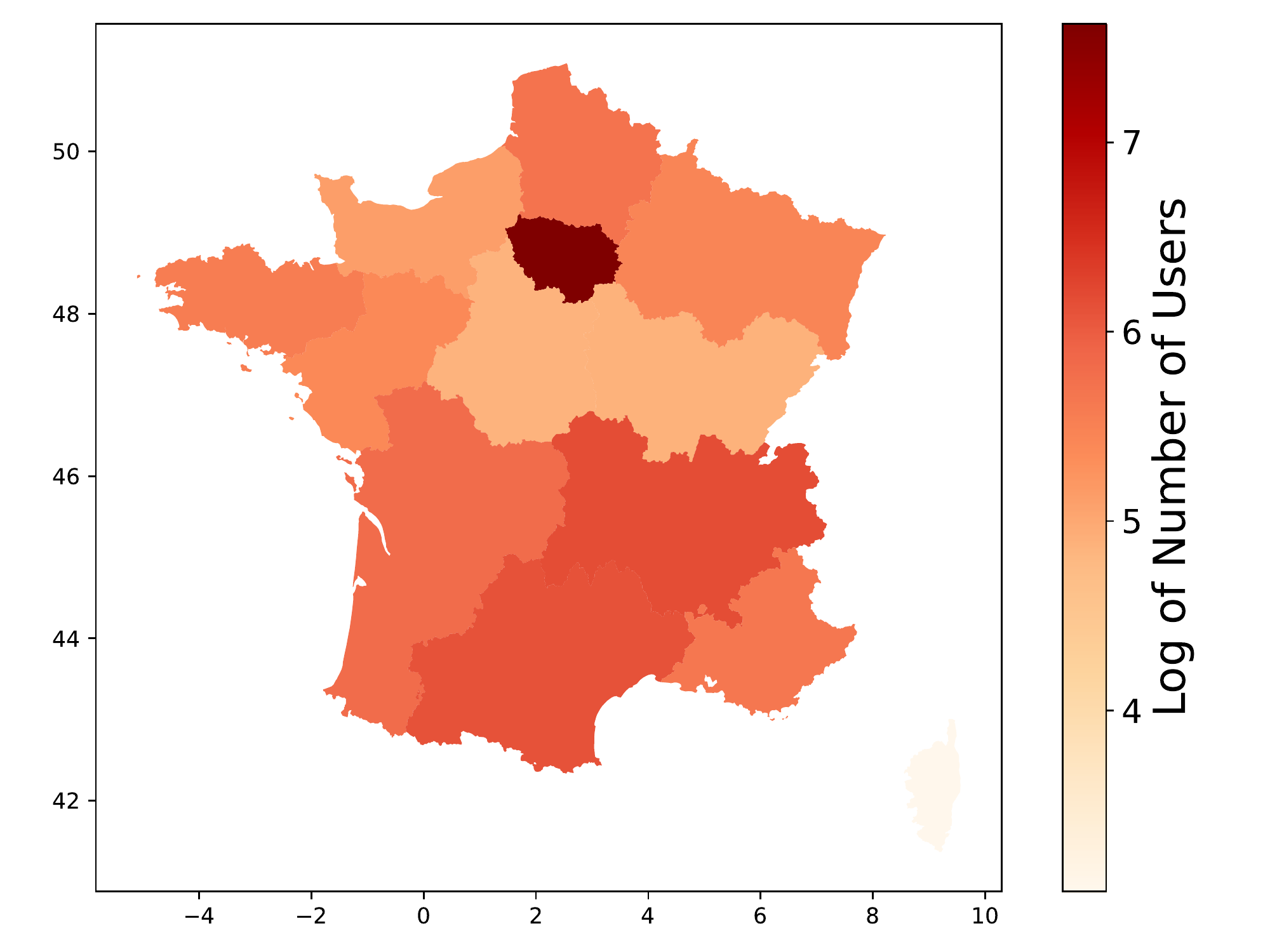}
	}\\
	\subfloat[Le Pen]{
	\includegraphics[width=0.45\textwidth]{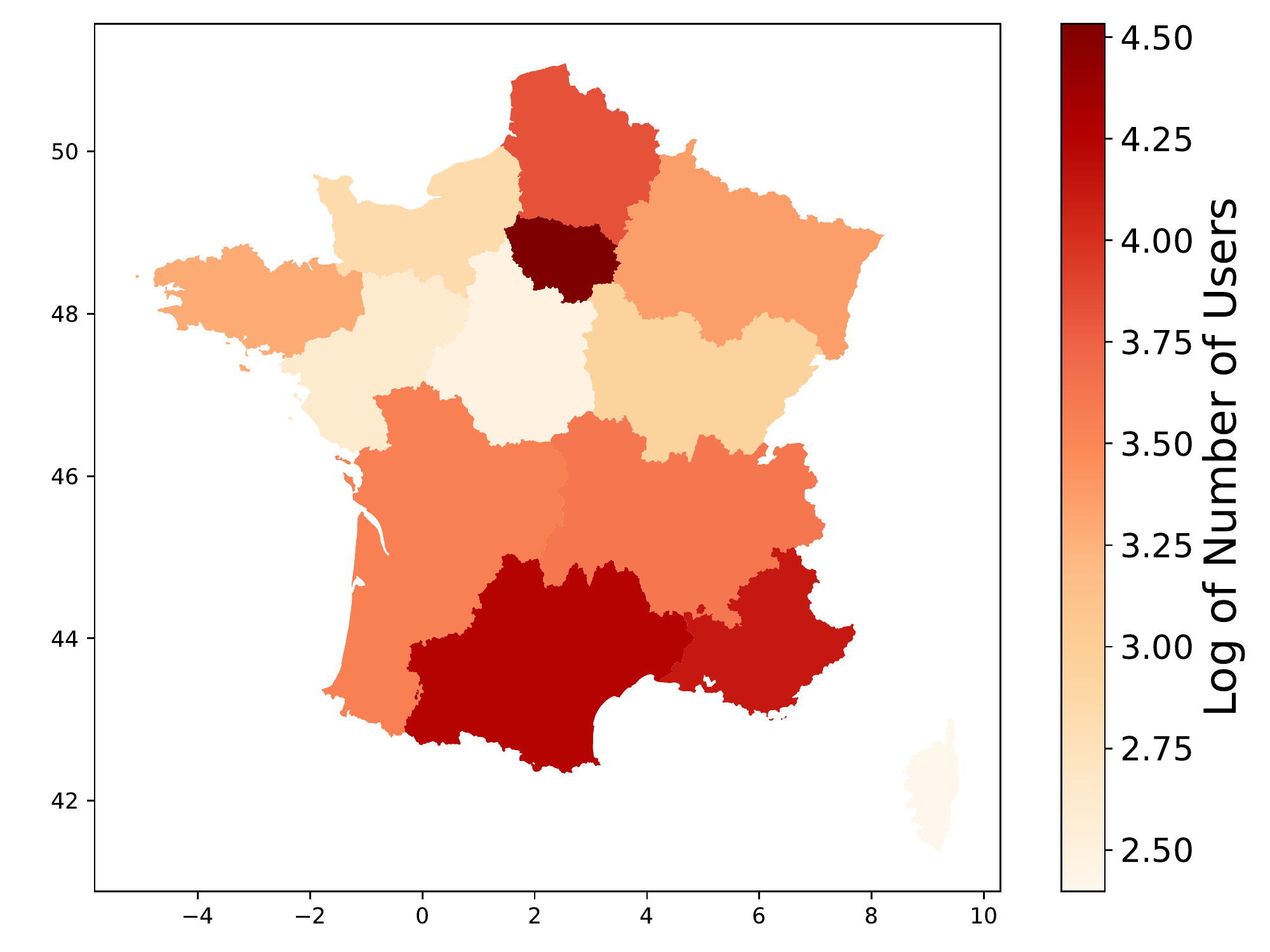}
	}
	\subfloat[Jadot]{
	\includegraphics[width=0.45\textwidth]{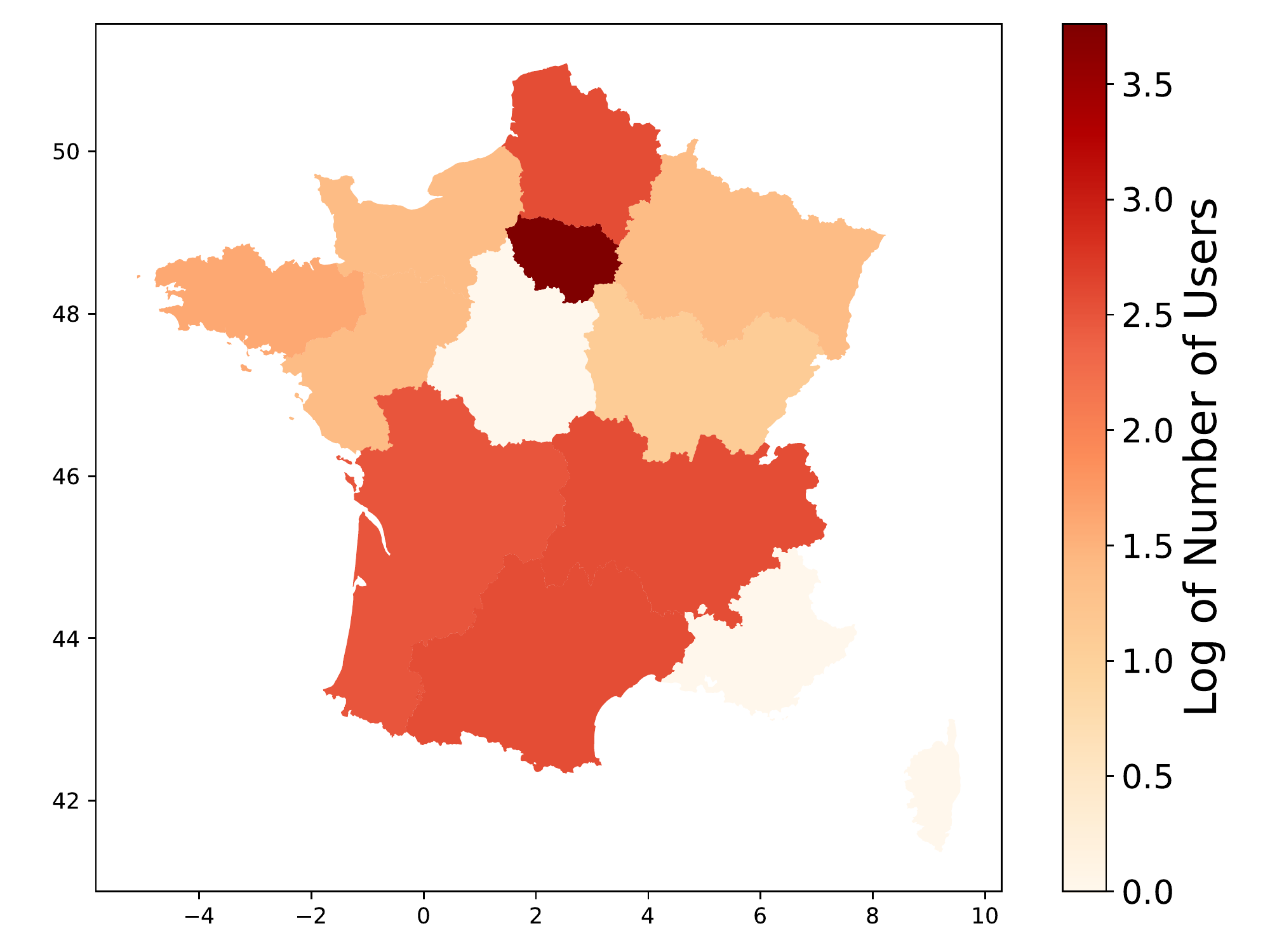}
	}\\
	\caption{Choropleth map of users in each community.}
	\label{geo}
\end{figure*}

\section{Dataset}
We create a novel Twitter dataset with the 2022 French presidential election as the central topic. The tweets used for building this dataset date from February 14, 2022 to April 5, 2022. This dataset corresponds to a large and coherent corpus consisting of small pieces of text related to the election candidates and major events during their campaigns. In our case, we focus on tweets that include tokens such as “présidentielle” and “élection”, as well as the names of candidates and their parties. 
We extract tweets in the French language containing the keywords mentioned above through Twitter's public streaming API. The public streaming API is able to extract a subset of the real-time Twitter stream. The resulting dataset consists of 62.6 million tweets and 1.2 million users.

\section{Graph-based Community Detection}

Given that the election has been at the center of attention in France since the beginning of 2022, daily campaigns, scandals, and debates are widespread through social media. We believe that there would be detectable online communities related to the different candidates and that each community will contain important information about each party's campaign. Therefore, we construct a retweet graph of Twitter users based on our dataset.

\begin{table*}[ht!]
\begin{center}

\begin{tabular}{||c | c | c ||}  
 \hline 
 Community & \# Accounts & Frequent hashtags \\ [0.5ex]  
 \hline\hline 
 Mélenchon & 15,001 & \#melenchonvagagner, \#melenchonsecondtour, \#jevotemelenchon \\  
 \hline 
 Anti-Macron & 12,428 & \#macrondehors, \#toutsaufmacron, \#macrondegage \\ 
 \hline 
 Zemmour & 12,101 & \#zemmourpresident, \#jevotezemmour, \#zemmourpresident2022  \\ 
 \hline 
 Macron & 5,816 & \#avecvous, \#macron2022, \#5ansdeplus \\ 
 \hline 
 Pécresse & 1,035 & \#valeriepresidente, \#pecresse2022, \#nouvellefrance\\ 
 \hline 

Le Pen & 1,001 & \#marinepresidente, \#dimanchejevotemarine, \#jevotemarine \\ 
 \hline 

Jadot & 196 & \#jadot2022, \#jevotejadot, \#totalsoutienàjadot\\ [1ex] 
 \hline 

\end{tabular} 
\caption{Statistics of different communities.}
\label{communities}
\end{center}
\end{table*}

\subsection{Graph Creation}
With our set of extracted tweets, their authors, and the users who retweet these tweets, we create a directed weighted graph $G = (V, E)$ where $n = |V|$ denotes the number of nodes. Specifically, each node represents a user on Twitter, and a weighted edge connects two nodes if one user retweets the other. For instance, the weight $A_{u, v}$ of the edge $(u, v)$ from vertex $v$ to vertex $u$ notes the number of times that the user $u$ retweeted the user $v$. We believe that a normal retweet indicates approval of the tweet's content, unlike the quote retweets and the replies. Therefore, the graph does not model the textual similarity of the tweets but only the relation between users. Note that the obtained graph G might contain self-loops that correspond to self-retweets. The obtained graph contains 1.2M nodes and 12.4M edges.

To avoid small, non-dense clusters formed from few numbers of users with few retweets, we decided to work on a dense subset of the graph that we obtained using the k-core decomposition algorithm instead of working on a full graph. The k-core decomposition algorithm \cite{kcore} aims to find subsets of a graph G. The subsets are called k-cores of G and are obtained by a recursive pruning strategy. Each node inside a k-core is connected to at least k other nodes inside this subset. The hyperparameter k is chosen so that we do not get a cluster with less than ten users. The final used k-core graph has 47,578 nodes and 8.2M edges. Seven communities are found in our dataset through applying this approach.

\subsection{Community Detection}

In graph mining, community detection helps to reveal the hidden relations among the nodes in a graph. Hence, to discern opinion groups inside the k-core graph, we apply the Louvain community detection method \cite{Blondel_2008} on an undirected version of our graph without self-loops. The Louvain method is chosen as it has reasonable computation costs while maximizing modularity. Moreover, it does not require fine-tuning of hyperparameters. It is therefore the only applicable method for finding communities in large graphs.

Seven communities were found after applying this approach to the user-based k-core graph. To get a general overview of the communities, we compute the frequencies of hashtags used by the users inside each community. We define the frequency of a hashtag as the number of users using this hashtag inside a community. We notice that first six out of these seven communities have remarkable hashtags with high frequencies that relate each community to a candidate in the election. For example, in one community with 16,190 users, the top three hashtags are \texttt{\#melenchonvagagner}, \texttt{\#melenchonsecondtour} and \texttt{\#jevotemelenchon} with respectively 16,073, 14,378 and 10,127 users. In contrast, no other hashtags supporting other candidates appear in this community's 50 most frequent hashtags. Thus, we label this community with "Mélenchon". Finally, the seventh and last community only holds keywords against the current President of France in the top 50 hashtags. We thus label it as the "Anti-Macron" community. In table \ref{communities} we detail the labels, the number of users, as well as the most biased and frequent hashtags for each community.

\subsection{Word Cloud Analysis}

To further analyze the political stances of the seven communities, we group the tweets posted by users of each community and generate a word cloud for each group of tweets. The resulting word clouds assemble the set of the most frequent unigrams and bigrams (after removing stop words) in each group of tweets, with their sizes proportional to their frequencies. Our analysis shows that except for the community with the "Anti-Macron" theme, all the other communities are distinctively composed of users who support one of the candidates among Emmanuel Macron, Valérie Pécresse, Éric Zemmour, Jean-Luc Mélenchon, Marine Le Pen and Yannick Jadot. The word clouds for each of these communities are shown in Figure \ref{wordcloud}. In addition to identifying which candidate each community of users supports, we can also use the word clouds to obtain a general idea of each candidate's campaign program. For example, the word cloud for Jadot's community contains words such as "justice social (social justice)" and "écologiste (ecologist)" with significant weights, and these concepts precisely lie at the center of his campaign program.

\subsection{Geolocation Analysis}

For each community of users, we plot the distribution of their geolocations within different regions of Metropolitan France. We select users with geolocation information and feed the declared locations into the geolocator from geopy\footnote{\url{https://github.com/geopy/geopy}} to obtain the region each area belongs to. We only consider users who are located within Metropolitan France. Although this limits our analysis to a subset of users from the dataset, the distribution can still reflect the overall situation. 

Given that users are remarkably concentrated in the Île de France region for all the communities, we plot the choropleth map with the number of users scaled by logarithm to visualize the variations for the other areas more clearly. The choropleth map is shown in Figure \ref{geo}. Through observing these maps, we can gain insights into each candidate's respective heartland. For instance, Le Pen has a much higher proportion of supporters in Occitanie, Provence-Alpes-Côte d'Azur and Hauts-de-France than all the other candidates. We also find that the supporters of Macron, Pécresse and Mélenchon are more evenly distributed among different regions outside of Île de France than the other three candidates. Zemmour and Le Pen both lack Twitter supporters in the middle and northwestern areas, while Jadot lacks supporters in the whole northern half of France except the region Hauts-de-France.

Our study was carried out right before the first turn of the election. We later did a follow-up of the actual results and found that the choropleth map of votes\footnote{\url{https://tinyurl.com/ycxya5dx}} bear a striking resemblance to our choropleth maps of communities.  

\begin{table*}[ht!]
\begin{center}

\begin{tabular}{||c | c | c | c ||}  
 \hline 
 Community & \# Unique Tweets & \# Offensive Tweets & Proportion of Offensive Tweets \\ [0.5ex]  
 \hline\hline 
 Mélenchon & 756,318 & 208,178 & 0.275 \\  
 \hline 
 Anti-Macron & 549,138 & 168,685 & \textbf{0.307} \\ 
 \hline 
 Zemmour & 1,034,538 & 316,214 & \textbf{0.305}  \\ 
 \hline 
 Macron & 468,138 & 126,122 & 0.269 \\ 
 \hline 
 Pécresse & 80,365 & 19,487 & 0.242 \\ 
 \hline 
Le Pen & 86,272 & 25,368 & 0.294 \\ 
 \hline 
Jadot & 12,340 & 1,632 & 0.132 \\ [1ex] 
 \hline 

\end{tabular} 
\caption{Offensive Tweets in Each Community}
\label{offensive}
\end{center}
\end{table*}

\section{Detection of Offensive Tweets}
Along with the growing popularity of social media and online platforms, the use of offensive online language has become a significant problem. Under such circumstances, automatic detection of offensive language has received much research attention \cite{risch-etal-2020-offensive}. With presidential elections being such a controversial topic, relevant tweets are bound to contain offensive language. It is expected that online supporters of a given candidate would make offensive comments towards other candidates and their supporters. However, supporters of the more extreme candidates might also be more inclined to use offensive language. The targets of offensive tweets might include other public groups in addition to opposing candidates and their supporters. We build an automatic classification model to detect offensive tweets in each political community and eventually compare the results across communities.

\subsection{Detection Model}
We initialize our model using BERTweetFR \cite{guo-etal-2021-bertweetfr} and fine-tune it on the MLMA Hate Speech Dataset \cite{ousidhoum-etal-2019-multilingual}. BERTweetFR is a French RoBERTa model \cite{liu2019roberta}, initialized using the general-domain French language model CamemBERT \cite{martin-etal-2020-camembert} and further fine-tuned on 16GB of French tweets. It achieves the state-of-the-art performance on French Twitter tasks. The MLMA Hate Speech Dataset is a multilingual multi-aspect Twitter dataset for hate speech analysis. We take a subset of this dataset selecting only French tweets labeled as either "normal" or "offensive". The resulting subset contains 821 normal tweets and 1690 offensive tweets. After fine-tuning for 3 epochs, our classification model achieves a f1 score of $83.96\%$ on a 80/20 train-test split.

\subsection{Detection Results}
We run our classification model for tweets posted by users in each political community. We only consider unique tweets, discarding retweets by deduplicating them based on the text content. This choice is made because we aim to detect the origination of offensive language rather than to analyze its propagation pattern. The detection results are shown in Table \ref{offensive}. A key observation is that users from the Anti-Macron and Zemmour communities are the most likely to post offensive tweets, reaching respective proportions of 0.307 and 0.305. This is in line with our expectations: the Anti-Macron community is naturally supposed to be more offensive as the main goal is to oppose and defy; as for Zemmour, he is a far-right candidate who has been personally fined \euro 10,000 for hate speech by a Paris court \footnote{\url{https://www.bbc.com/news/world-europe-60022996}}. We also observe that the communities of right-wing candidates tend to have more offensive content in general, with the only exception being Pécresse's who is a more moderate candidate. A possible explanation is the employment of automatic bots in her community. We will further elaborate on the topic of bots in the following section.

\section{Detection of Automatic Bots}

 It has come to light in recent years that a significant amount of Twitter accounts are controlled, at least partly, by software. Some research estimate that between 9\% to 15\% of all twitter accounts are somewhat automated \cite{DBLP:journals/corr/VarolFDMF17}. Bots are an important tool for opinion manipulation \cite{DBLP:journals/corr/SubrahmanianADK16}, and are being used to influence important subjects such as political elections (\cite{DBLP:journals/corr/Ferrara17aa},\cite{DBLP:journals/corr/abs-1902-00043}). Bots also help spread misinformation (\cite{DBLP:journals/corr/ShaoCVFM17}), and have impacted the online debate on vaccination \cite{doi:10.2105/AJPH.2018.304567}, with an estimated 45\% of COVID-19 related Twitter accounts exhibiting bot-like behavior (\cite{DBLP:journals/corr/abs-2008-00791}).
This section proposes to estimate the role bots are playing in the 2022 French election by comparing their relative use within each community.

\subsection{Detection model}
    
There is a multitude of available Twitter bot detection models (\cite{Lee_Eoff_Caverlee_2021}, \cite{DBLP:journals/corr/abs-1802-04289}, \cite{twitterspammer}, \cite{10.1145/3308560.3316504}), and APIs (\cite{DBLP:journals/corr/DavisVFFM16}) that use semantic, statistical or neighborhood properties to evaluate the likelihood of an account being a bot. However, due to the large amount of data we have to process, we need to use a scalable and generalizable models. This study is therefore going to rely on statistical features available in the user metadata object given by the Twitter API. Our employed model is similar to the one presented in \cite{DBLP:journals/corr/abs-1911-09179}, which is scalable and yields adequate generalization results on the task of bot detection \cite{DBLP:journals/corr/abs-2106-13088} in different scenarios.
    
    \subsubsection{Feature Selection}
    
        The list of available user metadata features relevant to bot detection is listed as below:
        
        \begin{itemize}
            \item \textsc{statuses count}
            \item \textsc{followers count}
            \item \textsc{friends count}
            \item \textsc{favourites count}
            \item \textsc{listed count}
            \item \textsc{default profile}
            \item \textsc{verified}
            \item \textsc{geographical location enabled}
        \end{itemize}
        
        These available features give other interesting statistical information that we compute as additional derived features for the model. Such features include the frequency of tweets (statuses count/user age), the respective growth rate of followers, friends, favorites and listed accounts (respective counts/user age). We also take into account information from the username, such as its length and the number of digits it contains. The length of user description is also proven to be a relevant feature \cite{DBLP:journals/corr/abs-1911-09179}.
        
        We choose random forest as our classifier, as it yields near-perfect results on any individually labeled dataset.
    
        \begin{table*}[ht!]
        \begin{center}
        \scalebox{0.82}{
            \tabcolsep=0.11cm
            \begin{tabular}{||c | c | c | c | c | c | c | c || } 
             \hline
             Community & \# Accounts & \# Bots & Proportion of bots in community & \# Tweets & \# Automated Tweets & proportion Automated tweets \\ [0.5ex] 
             \hline
             \hline
             Mélenchon & 15,001  & 2,507  & 0.167 & 5,755,664  & 1,273,656 & 0.284 \\
             \hline
             Anti-Macron & 12,428  & 2,181 & 0.175 & 5,435,820 & 1,268,240 & 0.304\\
              \hline
             Zemmour & 12,101  & 2,217 & 0.183 & \textbf{6,160,153} & 1,501,207  & 0.322  \\
             \hline
             Macron & 5,816   & 1,001 & 0.172 & 2,219,491 & 514,820   & 0.302 \\
             \hline
             Pécresse & 1,035 & 208 & \textbf{0.200} & 408,319 & 134,373  & \textbf{0.490}
             \\ 
             \hline
             Le Pen & 1001 & 184 & 0.184 & 463,290 & 127,633 & 0.380
             \\
             \hline
             Jadot & 196 & 30 & 0.153 & 66,541 & 19,876 & 0.426
             \\[0.5ex]
             \hline 
            \end{tabular}
            
            }
            \captionof{table}{Bots statistics of different communities}
        
        \label{Tab:BotStats}
        \end{center}
    \end{table*}
    
    \subsubsection{Training Data}
    
        The choice of training data for such a task is crucial. There are many different types of bots for different domains, and there is generally poor classification generalization across datasets \cite{DBLP:journals/corr/abs-1809-09684}. Considering the task at hand which is the classification of politically oriented Twitter users into bots or human labels, we decided to train our model on a concatenation of multiple available datasets: \textbf{Political-bots-2019} \cite{DBLP:journals/corr/abs-1901-00912} (a compendium of political bots), \textbf{midterm-2018} \cite{DBLP:journals/corr/abs-1911-09179} (a hand-labeled dataset of users and bots during the 2018 American midterm elections), \textbf{botwiki} \cite{DBLP:journals/corr/abs-1911-09179} (a collection of self identified Twitter bots),  \textbf{verified-2019} \cite{DBLP:journals/corr/abs-1911-09179} (a collection of verified Twitter users),  \textbf{Cresci 2019-2018} (\cite{DBLP:journals/corr/abs-1902-04506}, \cite{DBLP:journals/corr/abs-1804-04406}) (datasets of manually annotated bots), and finally \textbf{Twibot-20} \cite{DBLP:journals/corr/abs-2106-13088} (a comprehensive hand labeled dataset of Twitter bots). The statistics of each dataset is shown in Table \ref{data_bot}.

        \bgroup
        \def\arraystretch{1.5}
        \begin{center}
            
            \begin{tabular}{||c | c | c||} 
             \hline
             Datasets & \# human & \# bots \\ [0.5ex] 
             \hline
             \hline
             \textsc{Political-bots-2019} & 0 & 62 \\ 
             \hline
             \textsc{Midterm-2018} & 8,092 & 42,446  \\
             \hline
             \textsc{botwiki} & 0 & 698 \\
             \hline
             \textsc{verified-2019} & 1,987 & 0 \\
             \hline
             \textsc{twibot-20} & 5,237 & 6,589 \\
             \hline
             \textsc{cresci-18/19} & 6,514 & 7,455 \\
             \hline
             \textsc{total} & 21,830 & 57,250  \\ [1ex] 
             \hline
            \end{tabular}
        \captionof{table}{Statistics of training datasets.}
        \label{data_bot}
        \end{center}
        \egroup

    \subsubsection{Training results}
    
\begin{figure*}[ht]
            \centering
            \includegraphics[scale=0.45]{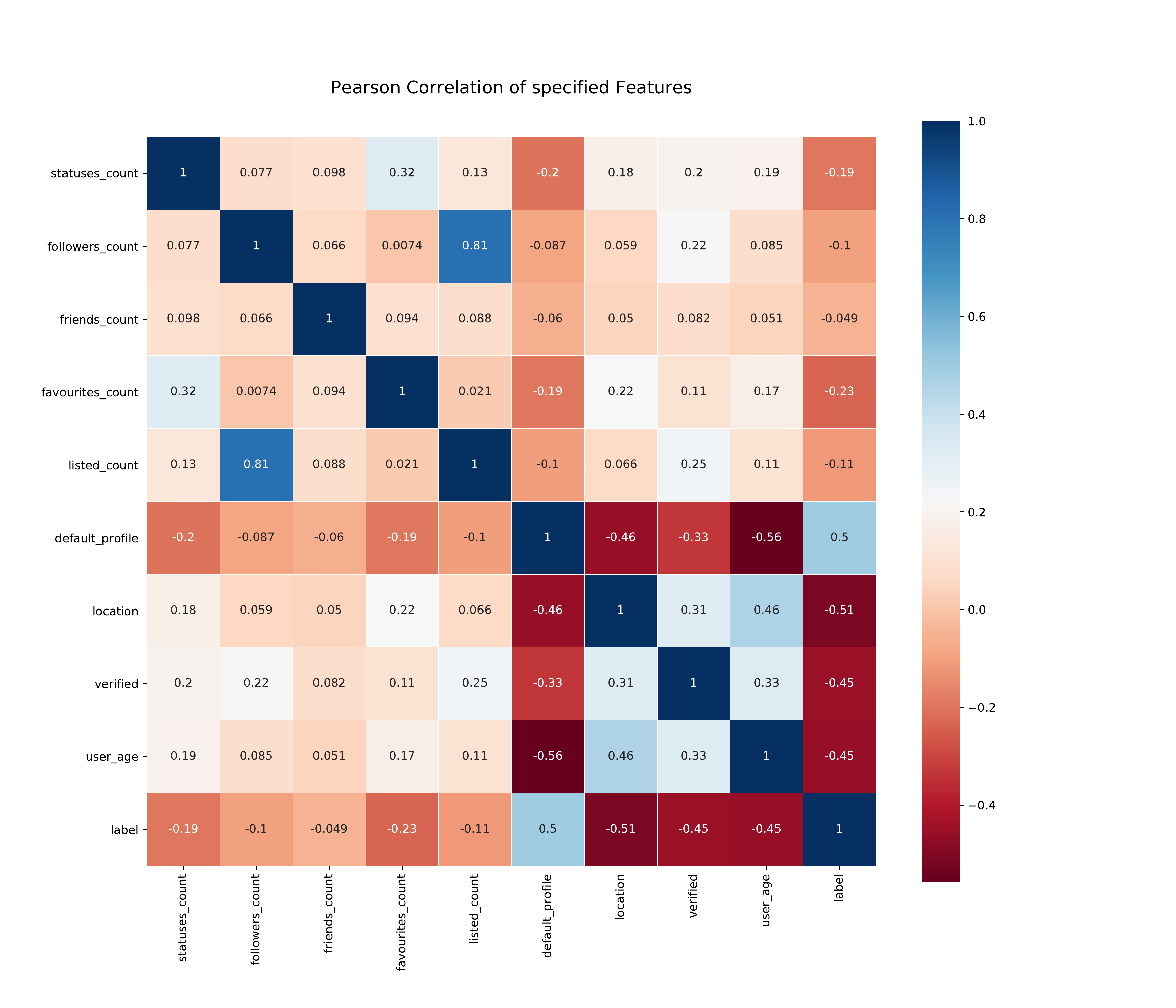}
            \caption{Heatmap of Pearson correlation between features and labels.)}
            \label{fig:Heatmap}
        \end{figure*}

        \paragraph{Correlation} 
            It is crucial to consider the most discriminative features when attempting a task like bot detection, the model should be interpretable. For example, as we see in Figure \ref{fig:Heatmap}, there is a strong correlation between the automation of an account and the age of the account, whether or not the account is verified or geolocalisation enabled. Another strong indicator of automation is the presence of a default-profile, which means an account with a lack of personalization (i.e. custom banner or profile picture).
            
        \paragraph{Feature importance} By studying the importance of each feature, we find that the number of statuses, the information of followers (such as its raw count and growth rate) and the age of the user are the most critical features for the classifier.
        
        \paragraph{Results} Our random forest classifier achieves a $95.0\%$ f1 score with 10-fold cross validation.
    
    \subsection{Bot Detection Results}
        
        From our experiments, we observe that there is a significant amount of automated accounts in our dataset -- with an estimation of at least 15\%, with a conservative labeling threshold, of accounts partaking in the debate coming from bots. We have tuned the classification threshold, which is usually at 50\% certainty to 75\%, considering the importance of precision in the case of bot classification.
        
        As shown in Table \ref{Tab:BotStats}, while the number of users in each community is significantly different, we find a similar proportion of bots for each cluster. Therefore, we conjecture that there has not been any large-scale operation to influence the election with bots from any side.
        
        Another insight we can get from Table \ref{Tab:BotStats} is the campaigning approach of each community. We do not deduplicate tweets to remove retweets, considering the importance of retweeting in automated accounts. For example, the cluster supporting Zemmour, while being smaller than some of the others, has significantly more tweets, including retweets per person, showing the particular engagement Zemmour supporters seem to offer online. Similarly, we can see that it is the only large cluster with the most considerable bot activity, albeit by a small margin.
        On the other side, we can see that the Pécresse cluster, smaller in scale, has heavy activities from bots. This difference in bot activity for the three smallest communities may come from factors such as some very dynamic automated news pages.

    \subsubsection{Limitations of Automatic Bot Detection}
        It is important to keep in mind that bot detection, while being effective, is a limited approach, especially in the case of political elections. While a lot of bots can be found, a nuance is to be made, as bots in a cluster are not necessarily promoting the candidate. Some of the more basic bots that promote cryptocurrency or fishing sites usually simply post the same messages repeatedly, along with all the popular hashtag at a time $t$. Such a behavior artificially inflates the number of bots we find in the community of candidates that are naturally more active on twitter. The same goes for "automatized behavior", as we see in figure \ref{fig:Heatmap} and in our feature importance section. While algorithms are accurate, their most discriminative features are the number of statuses, followers, and the age of the user. The situation can be more complicated in the case of politics, as there are actual people who are willing to tweet with the hashtag \#MélenchonPrésident one hundred times a day simply because they are extremely passionate about the campaign.

\section{Conclusion and Future Work}

In this paper, we have leveraged graph-based community detection methods to gather insights into each of the most significant candidates' online campaigns for the 2022 French presidential election. We have been able to build a portrait of the average voter for each candidate, the interest they carry in different political subjects, their geolocalization, and their language habits. We have also presented results on the usage of automated accounts, or the lack thereof, in each community.

Many future tasks are possible to be performed on this dataset. Based on the community detection of political communities, a relevant study could be to analyze of the impact of major political events or debates. Considering that we have collected tweets from February to April on a daily basis, we could quantify the shift in the communities after debates between two candidates or how the start of the Ukraine war influenced electors. In the same way, we could also investigate the shift between the two turn of votes. French elections are based on a two-turn system, with the first turn aiming at narrowing down the the list of candidates and only keeping the two largest ones. The continued gathering of data and community detection could show us which communities turn to which candidate during the period between the two turns and how their language habits evolve.



\section{Bibliographical References}\label{reference}
\bibliographystyle{lrec2022-bib}
\bibliography{arxivbib}

\end{document}